\documentstyle[twoside,fleqn,espcrc2]{article}

\def\a{\alpha}
\def\b{\beta}
\def\c{\chi}

\def\g{\gamma}
\def\h{\eta}

\def\k{\kappa}     

\def\m{\mu}
\def\hm{\hat{\mu}}

\def\p{\pi}       

\def\J{\Psi}

\def\Jb{\overline{\J}}
\def\cb{\overline{\c}}

\newcommand{\Vd}{V^{\dagger}}

\newcommand{\del}{\partial}
\newcommand{\av}[1]{\mbox{$\langle #1 \rangle$}}

\newcommand{\half}{\mbox{{\normalsize $\frac{1}{2}$}} }

\newcommand{\eigth}{\mbox{{\normalsize $\frac{1}{8}$}} }
\newcommand{\thtwo}{\mbox{{\small $\frac{1}{32}$}} }

\newcommand{\RE}{\mbox{Re\,}}
\newcommand{\IM}{\mbox{Im\,}}
\newcommand{\Tr}{\mbox{Tr\,}}
\newcommand{\ra}{\rightarrow}

\newcommand{\dg}{\dagger}

\newcommand{\aplt}{ \mbox{}_{\textstyle \sim}^{\textstyle < }     }
\newcommand{\apgt}{ \mbox{}_{\textstyle \sim}^{\textstyle > }     }

\def\be{\begin{equation}}
\def\ee{\end{equation}}
\def\bea{\begin{eqnarray}}
\def\eea{\end{eqnarray}}

\newcommand{\AmS}{{\protect\the\textfont2
  A\kern-.1667em\lower.5ex\hbox{M}\kern-.125emS}}

\title{ \vspace{-11mm}\hbox{}
         \hfill ITFA-92-31, UCSD/PTH 92-42\vspace{11mm} \\
       Staggered fermion approach to chiral gauge theories
       on the lattice \thanks{Presented by J.C. Vink}   }

\author{Wolfgang Bock \address{Institute of Theoretical Physics,
                              University of Amsterdam,
                              Valckenierstraat 65, 1018 XE Amsterdam,
                              The Netherlands},
        Jan Smit$^{\;\; {\rm a}}$
        and
        Jeroen C. Vink \address{University of California, San Diego,
              Department of Physics-0319,   La Jolla, CA 92093-0319, USA}
        }
\begin{document}
\begin{abstract}
The staggered fermion approach to build models with chiral fermions is briefly
reviewed. The method is  tested in a U(1) model with axial
vector coupling in two and four dimensions.
\end{abstract}

\maketitle

\section{INTRODUCTION}

An important open problem in lattice field theory is to find a
satisfactory method to describe chiral fermions. There exist several
attempts and proposals, for an overview see \cite{Petch92},
 but none of these has so far been shown to work. Here we shall
discuss and test an approach \cite{Smit88} based on the staggered
fermion method. Note that  in this approach the notorious
species doublers need not be decoupled, since they are used as
physical spin-flavor degrees of freedom.

Staggered fermion fields, denoted by $\c_x$, do not carry explicit flavor
and Dirac labels since these components are `spread out' over the
lattice. In the classical continuum limit one can see the usual Dirac
and flavor structure in momentum space \cite{GoSm87},  and an equivalent
method can also be given in position space. Thereto
one introduces $4\times4$ matrices $\J^{\a\k}_x$ defined as,
\be
   \J_x = \eigth\sum_b \g^{x+b}\c_{x+b},\;
   \Jb_x = \eigth\sum_b (\g^{x+b})^{\dg}\cb_{x+b},
\ee
with $\g^{x}=\g_1^{x_1}\cdots\g_4^{x_4}$ and the sum running over the
corners of a  hypercube, $b_{\m}=0,1$.  Note that these
$\J_x$ are not independent, only their Fourier components in the
restricted momentum range $\pi/2 < p_{\m} \leq \pi/2$ may be
considered independent.
The index $\a(\k)$ on $\J$ acts like a Dirac (flavor) index and one can
construct models involving arbitrary spin-flavor couplings to Higgs or
gauge fields in a straightforward manner
\cite{Smit88}.  In the classical continuum limit (for
smooth external fields and for the low momentum modes of $\J$)
the desired symmetry properties of the model emerge. The
important issue is  if the same can be achieved when the
effect of quantum fluctuations is taken into account.

This approach has been successfully applied to a study of
a fermion-Higgs model \cite{BoSm92}.
Here  we consider a
U(1) model with axial-vector coupling in two and four dimensions,
defined by the action in the $\J$ notation, $(d=4)$,
\bea
 S_F & = & -\sum_{x\m} \half\Tr \left[
      \Jb_x \g_{\m}(U_{\m x}P_L + U^*_{\m x}P_R)\J_{x+\hm}
                 \right. \nonumber\\
   & - &  \left.\Jb_{x+\hm} \g_{\m}(U^*_{\m x}P_L + U_{\m x}P_R)\J_{x} \right].
                   \label{MODPSI}
\eea

By working out the trace in ({\ref{MODPSI}) one finds the action for the
staggered field,
\bea
 S_F  = - \thtwo\sum_{x\m}
           \left[ (\h_{\m x}\sum_b c^{\m}_{x-b} \cb_x\c_{x+\hm} - h.c.)
                    \right. & & \nonumber\\
  +  \left. (i\h_{5x}\h_{\m x+n}  \sum_{b+c=n}  s^{\m}_{x-b}
         \cb_x\c_{x+c-b+\hm} + h.c.) \right] & &   \label{MODCHI}
\eea
where $c^{\m}_x=\RE U_{\m x}$ and $s^{\m}_x=\IM U_{\m x}$. The
sign factor $\h_{\m x}=(-1)^{x_1+\cdots +x_{\m-1}}$ represents the Dirac
$\g_{\m}$,  $\h_{5x}=\h_{1 x}\h_{2 x+\hat{1}} \cdots
\h_{4 x+\hat{1} + \cdots +\hat{4}} = (-1)^{x_1+x_3}$ represents
$\g_5$ and $n=(1,1,1,1)$.
In the classical continuum limit this action describes four (two)
flavors of axially coupled Dirac fermions in four (two) dimensions.

In the $\J$ form the action appears to be gauge invariant
under $U \ra U^g$, $\J \ra \J^g$, with
\bea
  U^g_{\m x} &=& g_x U_{\m x} g^*_{x+\hm}, \;\; g_x \in U(1),
\label{UG}\\
\J_x^g &=& (g_x P_L + g^*_x P_R)\J_x. \label{PSITRAN}
\eea
However, since the $\J_x$ are not independent, such a local
transformation cannot be carried over to the $\c$ field, and the
action is not gauge invariant.
\section{GAUGE (NON)INVARIANCE}
Even though the action (\ref{MODCHI}) is not gauge invariant, it is
well-known \cite{Smit88} that the model
is closely related to a gauge invariant model.
Let $S= S_F + S_{ct}$ be the gauge non-invariant action, consisting
of the term (\ref{MODCHI}) plus additional counterterms, denoted by
$S_{ct}$.
The partition function can be written as,
\bea
 Z  & = &   \int D\c' D\cb' DU' e^{S(U',\c',\cb')} \nonumber \\
  & \stackrel{U'= U^{\Vd}}{=} & \int D\c' D\cb' DV DU
e^{S(U^{\Vd},\c',\cb')}.
\eea
Here we have used invariance of the Haar measure  $DU=DU^g$,
$g=\Vd$, and $\int DV=1$.
By explicitly writing the `longitudinal gauge mode' $V$, the
action $S(U^{V^\dg},\c,\cb)$ is seen to be gauge invariant under
(\ref{UG}) and $V_x\ra g_x V_x $.

In the scaling region this  gauge mode $V$ may decouple from
the physical states (if the fermion content is anomaly free),
which is to be achieved by choosing the appropriate counterterms
in $S_{ct}$. In particular  we add a mass term for the gauge
field,
\bea
 S_{ct} &=& \k \sum_{x\m} (U'_{\m x} + h.c.) , \label{SKAP}\\
        &=& \k \sum_{x\m} (V^*_x U_{\m x} V_{x + \hm} + h.c.).
                                   \label{HKIN}
\eea
The $V$ field can also be viewed as a radially frozen Higgs
field. Then (\ref{HKIN}) acts like a kinetic term for the Higgs
field. One expects a critical line $\kappa=\kappa_c(\b)$ (with $\b$
the bare gauge coupling) and for $\k\ra \k_c$ the Higgs mode will
be present as a physical state. By keeping $\k$ away from $\k_c$
in the symmetric
phase, the scalar mass will be of the order of the cut-off
and $V$ will decouple.

The attractive feature here is that gauge fixing is avoided.
Alternatively one might  try to fix the
gauge
and then remove undesired couplings of the $V$ field by adding
suitable counterterms, as in the `Rome approach' \cite{Rome89}.
One then has to worry about technical and Gribov problems with
non-perturbative gauge fixing.

Using $V$ as a Higgs field, we may think of $U'$ as the gauge
field in the unitary gauge, and similar for $\J'$ and $\c'$. It
is natural to add a fermion mass term
\be
S_Y = -y \sum_x\Tr\Jb'_x\J'_x, \label{YUK}
\ee
which would turn into a Yukawa coupling $-y\sum_x\Tr\Jb_x(V_x^2
P_R+(V^*_x)^2 P_L)\J_x$ if $\J'_x = (V^*_x P_L + V_x P_R) \J_x$
could be implemented as a transformation of variables.
\section{EXTERNAL GAUGE FIELDS}
 We shall first test our model for smooth external gauge
fields, i.e. $U\approx 1$ everywhere, and $V=1$, in two dimensions.
Since our model has axial-vector coupling, the vector current
has an anomaly. One therefore expects the relation,
\be
  \av{\del'_{\m} J_{\m x} }_{\c} = \frac{1}{2\p} F_x,
                          \label{CURDIV}
\ee
where the notation $\av{\bullet}_{\c}$ means integration over the $\c$
field.
The  vector current $J_{\mu}$ (given in $\J$ notation for simplicity)
and field strength $F$ are defined as,
\bea
  J^V_{\m x} & = & i\Tr (\Jb_x\g_{\m}U_{\m x}\J_{x+\hm} + h.c.),
                                 \label{VECCUR} \\
  F_{x+n/2} & = &  -i\log U_{1x}U_{2 x+\hat{1}}U^*_{1x+\hat{2}}U^*_{2x},
\eea

As an example we consider a spatially constant $U$ field defined by,
\be
U_{1 x} = \exp(iA_1 \sin (2\p t/L)),\; U_{2x} = 1.
                    \label{USMOOTH}
\ee
 Lattice sites are
denoted by $x\equiv (z,t)$ and the lattice size is $L^2$.
Fig. 1 shows the divergence of the current (the left hand side of eq.
(\ref{CURDIV}), denoted by $\Box$) and the anomaly (full line).
Clearly the correspondence is excellent and this
remains the case for amplitudes $A_1\aplt 0.2$. For larger values one
begins to see deviations.
\begin{figure}[htb]
\vspace{60mm}
\caption{Current divergence ($\Box$) and anomaly (full line) for the
smooth  gauge field (\protect{\ref{USMOOTH}}), $L=24$.
The ($\ast$) includes averaging
over $V$ field fluctuations with $\protect{\k}=0.6$, $L=8$.
}
\label{fig:1}
\end{figure}

Our second test case uses an external gauge field with non-trivial
topology. For the field  (\ref{USMOOTH}) the topological
charge $Q=\sum_x F_x/2\p$
is zero.
We can construct a gauge field with non-zero $Q$ and constant field strength
$F_x = 2\p Q/L^2$, by writing
\be
\begin{array}{ll}
U_{1 x} = \exp(iFt),  & U_{2x}   =  1\;  ,t=1,\cdots,L-1; \nonumber\\
                     & U_{2x}    =  \exp(iFLz),\; t=L.
                            \label{UTOP}
\end{array}
\ee
For small $Q/L$, the link field $U$ is close to one everywhere
{\em except} on the time
slice $t=L$ which contains  a transition function.
     This transition function makes $U$ `rough' (discontinuous) at $t=L$.
With a gauge invariant model such a transition function is invisible, but
since our action and the current (\ref{VECCUR}) lack gauge invariance we
may expect deviations from the anomaly equation (\ref{CURDIV}) at this
timeslice. This is illustrated in fig. 2, which is the same as fig. 1 but
now for the gauge field (\ref{UTOP}).
The presence of the slice with the transition function is clearly
visible and its disturbance extends over roughly 6 time slices. We have
checked that on a larger lattice the disturbance remains confined to
this number of time slices and one might argue that in the infinite
volume limit the effect of the transition function becomes an invisible
surface effect.
\begin{figure}[htb]
\vspace{60mm}
\caption{ Same as fig. 1, now for the gauge field (\protect{\ref{UTOP}})
with $Q=1$ and without $V$ field fluctuations. }
\label{fig:2 }
\end{figure}

These results show that for smooth fields the correct anomaly structure
is reproduced by the staggered fermions. Also non-trivial topological
sectors can presumably be incorporated in the proper way, but the strong
effect of the roughness produced by the transition functions hints at
difficulties for fields subject to quantum fluctuations.
\section{FLUCTUATING $V$ FIELD}
Suppose we decompose the gauge field in a `transverse' and a
`longitudinal' part,  $U'_{\m x}=V_x U_{\m x}^{tr}V^*_{x+\m}$, where
$U^{tr}$ satisfies the (lattice) Landau gauge condition. The
fluctuations of $U^{tr}$ are governed by a gauge coupling $\b$ and
 get suppressed for large $\b$.
If not used as a physical Higgs field, the $V$ modes should
be unobservable in the low energy physics, if the model becomes
gauge invariant. This can be tested already for $U^{tr}\ra 1$.

We have investigated the resulting fermion-scalar model
for $y\neq 0$. For $y\ra 0$ and $\k$ strictly smaller than $\k_c$
the $V$ field should decouple. In the quenched approximation
we  have computed the fermion propagators
 $\langle \J_x \Jb_y\rangle$ and $\langle \J'_x \Jb'_y\rangle$,
where $\J_x \equiv (V_x P_R + V^*_x P_L)\J'_x$. Only the
first propagator should then show a physical particle pole (in
the symmetric phase). We found encouraging results in the broken
phase but ambiguous results, hampered by huge statistical
fluctuations, in the symmetric phase.

As an alternative test  we add the longitudinal modes
to the smooth external gauge field (\ref{USMOOTH}),
$U_{\m x} \ra U'_{\mu x}=V_x U_{\m x}V^*_{x+\m}$, and compute
the anomalous divergence relation (\ref{CURDIV}) with the path integral
average over $V$ included ($y=0$).
For large $\k\apgt 1$ we then reproduce  the result for the
fixed background, up to statistical errors.
For smaller $\k$, approaching the
phase transition at $\k_c\approx 0.5$ we find that the current
gets renormalized.  The result in fig. 1 (denoted by $\ast$) shows that
the current strength is reduced by a factor $\approx 0.7$ at $\k=0.6$.
Such a  renormalization is to be expected for staggered currents
and has been studied previously in QCD, see e.g. \cite{SmVi88}.

For smaller values of $\k$, in particular inside the symmetric phase, the
magnitude of the statistical fluctuations grows dramatically and
prohibits computation of the average current divergence.
\begin{figure}[t]
\vspace{60mm}
\caption{Possible phase diagrams in the $\k-y$ plane of the four dimensional
axial U(1) model. PM (FM) indicates the symmetric (broken) phase, the PM
phase at small $y$ is absent if the solid lines are true.}
\label{fig:3}
\end{figure}
\section{DYNAMICAL FERMIONS}
Since the back reaction of the fermions on the $V$ field should reduce
the troublesome fluctuations, we have also studied the
scalar-fermion model
 in four dimensions with dynamical fermions.
Here we focus on the phase diagram of the model in the $y-\k$
plane. For large $y$ one can use a hopping
expansion to see that the fermion field $\J'$ in this unitary gauge has
a pole whereas the fermion field
 $\J$ does not.
If this persists also for small $y$ the model fails. If the model
is viable we must see a cross over at some (small) value of $y$.

Unfortunately the  problems which hampered the quenched computations
appear to haunt us here as well: For small values of $y\aplt 0.4$ the
model has extremely long living metastable states, reminiscent of
spin glass dynamics and we have so far not been able to compute the phase
structure in the region $y\aplt 0.4$. The phase diagram must be
symmetric for $\k \ra -\k$, which we can use to check our results.
In fig. 3 we show the measured
phase transitions for large $y$ and two speculations for the structure at
small $y$. One possibility, the solid lines, represent first order
transitions and would rule out our model. A second possibility is indicated
by the dashed, second order lines and in this scenario our model is
still alive.
\\[2mm]
The numerical calculations were performed on the CRAY Y-MP4/464
at SARA, Amsterdam.
This research was supported by the ``Stichting voor
Fun\-da\-men\-teel On\-der\-zoek der Materie (FOM)''
and by the ``Stichting Nationale Computer Faciliteiten (NCF)''.
%
%
%

\end{document}